# Self-Limiting Mechanism of Anti-Stokes Optical Cooling in Diamond NV Centers

Haruki Manaka[1] and Yasuhiro Yamada[1,*]

[1]Graduate School of Science, Chiba University, Chiba, Inage 263-8522, Japan

**ABSTRACT**. Anti-Stokes optical cooling in diamond nitrogen-vacancy (NV) centers is experimentally and numerically investigated. Photoluminescence-excitation spectroscopy reveals pronounced phonon-assisted anti-Stokes emission under excitation below the zero-phonon line (ZPL). However, the below-ZPL excitation drives photoinduced charge-state conversion between negatively-charged $NV^-$ and neutral $NV^0$, thereby suppressing the $NV^-$-mediated cooling channel. Time-resolved photoluminescence (PL) measurements reveal an increase in the effective PL lifetime with excitation density, reflecting an increasing $NV^0$ contribution. By fitting nanosecond and millisecond PL dynamics with a minimal rate-equation model, we extract effective optical pumping and charge-conversion rates, which enables us to quantitatively simulate the cooling performance. The simulations predict a self-limiting behavior of anti-Stokes cooling and clarify the excitation conditions under which net cooling can be sustained within this effective model. The estimated cooling power per NV center is comparable, on a microscopic basis, to values discussed for semiconductor quantum dots and rare-earth optical coolers. These results identify charge-state conversion as a key bottleneck for defect-based optical refrigeration.

## I. INTRODUCTION

Anti-Stokes optical cooling is a technique that removes thermal energy from a material through photoluminescence (PL) with photon energies higher than that of the excitation light. When the external quantum efficiency (EQE) of anti-Stokes PL is sufficiently high, the absorption–emission cycle results in a net loss of internal energy, leading to cooling of the system. Since the theoretical proposal by Pringsheim in 1929 [1], anti-Stokes optical cooling has been investigated for many decades. Experimental demonstrations have been reported in rare-earth–doped solids [2-4] as well as in semiconductor nanostructures [5-8].

Despite this progress, anti-Stokes optical cooling in existing material systems is subject to intrinsic limitations. In rare-earth-based materials, near-unity internal PL quantum efficiency can be achieved, but the weak absorption cross section of 4f transitions limits the attainable cooling power [3,9]. In semiconductor nanostructures, by contrast, strong absorption enables high-density excitation, yet Auger recombination under such conditions imposes a fundamental limit on the achievable cooling performance [8,10]. In addition, photoinduced degradation and associated loss of PL efficiency often become non-negligible under the excitation conditions relevant to cooling [11,12]. Thus, the factors limiting optical cooling differ substantially among material platforms.

Against this background, nitrogen-vacancy (NV) centers in diamond have emerged as an alternative platform for anti-Stokes optical cooling [9,13,14]. NV centers exhibit high PL efficiency and exceptional photostability [15,16], and phonon-assisted anti-Stokes PL has been experimentally reported, suggesting their potential applicability to optical cooling [17,18]. Furthermore, as isolated point defects in a wide-bandgap host, NV centers are inherently free from multiexciton formation and Auger recombination processes that limit cooling in semiconductor quantum dots [8], representing a significant advantage from the perspective of nonradiative loss suppression. Diamond is also biocompatible and nontoxic, opening prospective bio-related applications of optical cooling, for example in cryosurgery [13,19].

However, most previous studies on NV centers have focused on the observation of anti-Stokes PL and its spectroscopic characteristics, and the feasibility of steady-state anti-Stokes optical cooling using NV centers has not yet been sufficiently examined. In particular, NV centers are known to undergo photoinduced charge-state conversion between the neutral ($NV^0$) and negatively charged ($NV^-$) states [18,20,21], yet the impact of these charge-state dynamics on the conditions required for optical cooling has not been systematically clarified.

In this work, we combine anti-Stokes PL spectroscopy, time-resolved PL measurements, and numerical simulations to quantitatively assess both the potential and the limitations of optical cooling in diamond NV centers. We show that, unlike semiconductor nanostructures where Auger recombination limits cooling under intense excitation, NV centers exhibit a distinct self-limiting mechanism: photoinduced charge-state conversion depletes the NV⁻ population that mediates the cooling cycle. Our results identify charge-state conversion as a key bottleneck for defect-based optical cooling and provide experimentally grounded guidelines for evaluating its feasibility.

*Contact author: yasuyamada@chiba-u.jp

## II. METHODS

Single-crystal diamond containing NV centers was purchased from Thorlabs (3.0 × 3.0 × 0.5 mm³, {100} surface). The sample contained approximately 4.5 ppm of NV centers, as specified by the supplier.

Photoluminescence excitation (PLE) spectroscopy was performed using a picosecond pulsed supercontinuum light source (repetition rate 3 MHz, seed pulse width < 100 ps) spectrally selected using a monochromator. The excitation photon energy was scanned in the range of 1.67–2.75 eV, covering the zero-phonon line and its low-energy side. The excitation linewidth was approximately 3 meV, and the average excitation power was 10 μW. The excitation beam was focused onto the sample to a spot diameter of approximately 15 μm. Spectra covering both NV⁻ and NV⁰ PL bands were recorded using a CCD spectrometer. The spectra were calibrated for the system spectral response and normalized by the excitation density. The excitation density used for the PLE measurements was sufficiently low that photoinduced charge-state conversion was negligible.

Nanosecond-scale PL lifetimes were measured using time-correlated single-photon counting. Spectrally selected PL was detected using a single-photon avalanche photodiode. The excitation source was a nanosecond pulsed supercontinuum light source (repetition rate 20 kHz), spectrally filtered to the desired wavelength.

Millisecond-scale PL dynamics were measured by modulating a 1.95-eV laser diode using an optical chopper. The PL signal was detected by an amplified photodetector and recorded using a digital oscilloscope.

Temperature-dependent measurements in the range of 280–320 K were performed using a temperature-controlled microscope stage. Unless otherwise noted, all data were acquired at room temperature.

## III. RESULTS

### A. Anti-Stokes PL and Optical Cooling Gain

To assess the feasibility of anti-Stokes optical cooling, it is essential to characterize the anti-Stokes PL properties in detail. Figure 1(a) shows representative PL spectra obtained under excitation at different photon energies. Under excitation below 2.1 eV, the PL is dominated by the negatively charged NV⁻,

*Contact author: yasuyamada@chiba-u.jp

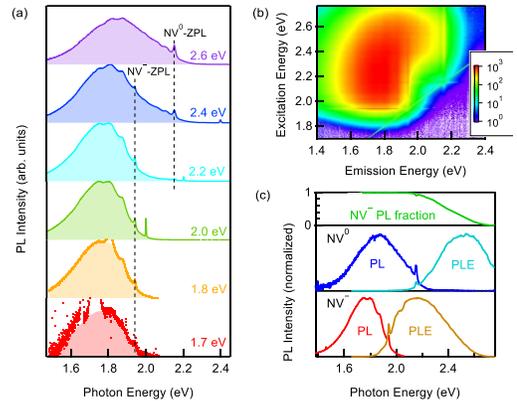

FIG. 1 (a) PL spectra measured under different excitation photon energies. (b) PLE contour map. (c) PL and PLE spectra for NV⁻ (lower panel) and NV⁰ (middle panel). NV⁻ and NV⁰ PL spectra were obtained under excitation at 1.74 and 2.76 eV. The PLE spectra were extracted by fitting the measured PL spectra and decomposing them into NV⁻ and NV⁰ spectral components. Top panel shows the fraction of the NV⁻ PL component as a function of excitation photon energy.

exhibiting a zero-phonon line (ZPL) at 1.94 eV together with several broad phonon sidebands (PSBs) on the low-energy side, while no contribution from the neutral NV⁰ is observed in this spectral window. Distinct PSB features appear at energy separations of approximately 65 meV and 135 meV from the ZPL. These features are attributed to vibronic emission involving localized vibrational modes of the NV center, with energies of approximately 64 meV and 138 meV, respectively, consistent with previous reports [22,23]. Notably, the overall spectral shape remains unchanged upon varying the excitation photon energy. Thus, excitation below the ZPL gives rise to anti-Stokes PL. The anti-Stokes PL observed here is not attributed to a two-photon excitation process, as confirmed by its linear dependence on the excitation density (see Supplemental Material) [24].

To provide a more comprehensive view of the optical response, Fig. 1(b) presents a PLE map acquired over a wider detection energy range, in which PL from both NV⁰ and NV⁻ can be observed. Under excitation above 2.16 eV, NV⁰ state is also photoexcited, as seen from the PLE map. The clear separation between the NV⁰- and NV⁻-related PL bands enables independent identification of their optical contributions under different excitation conditions. Fig. 1(c) shows the NV⁻ fraction as a function of excitation photon energy, which is obtained by spectral decomposition.

As further illustrated by the PLE spectrum in Fig. 1(c), the PL and PLE spectra exhibit an approximate mirror

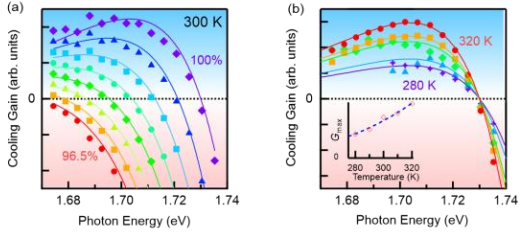

FIG. 2 (a) Cooling gain spectra at 300 K assuming EQE values from 96.5 to 100% in steps of 0.5%. (b) Cooling gain at different temperature from 280 to 320 K in steps of 10 K. EQE is assumed to be 100%. The inset shows the maximum cooling gain as a function of temperature. The dashed curve represents a fit to the Arrhenius relation ($G_{\max} \propto \exp(-\Delta E/k_\mathrm{B}T)$), where $\Delta E = 156$ meV. Solid curves are guides to the eye.

symmetry, which is characteristic of optical transitions in localized electronic systems. These observations confirm that the observed anti-Stokes PL originates from phonon-assisted transitions of localized NV centers rather than from band-like states.

Based on these spectroscopic characteristics, we evaluate the potential cooling performance using a simple energy-balance model. Here, for simplicity, we neglect the photoinduced charge-state conversion, which will be discussed below. To this end, we introduce the cooling gain, $G$, defined as the net energy extracted from the system per incident photon,

$$G(E_\mathrm{ex}) = A(E_\mathrm{ex})\{\eta_\mathrm{ext}\langle E_\mathrm{PL}\rangle - E_\mathrm{ex}\}, \quad (1)$$

where $A$ is the absorption probability under excitation at $E_\mathrm{ex}$, $\langle E_\mathrm{PL}\rangle$ is the average PL photon energy, and $\eta_\mathrm{ext}$ is the EQE. Figure 2(a) shows the calculated cooling gain spectrum at 300 K. Assuming unity EQE, a crossover from cooling to heating occurs near the PL peak energy (~1.73 eV). As the EQE decreases, the breakeven excitation energy shifts toward lower photon energies.

This analysis indicates that anti-Stokes optical cooling requires an exceptionally high EQE, approximately 97% or higher. Achieving such a high efficiency in bulk diamond is experimentally challenging. Indeed, the EQE of the present sample, as measured using an integrating sphere, is below 5% [24]. Nevertheless, previous studies have reported internal quantum efficiencies exceeding 80% for NV centers [33]. In addition, single-emitter studies have shown that the intrinsic quantum efficiency of NV centers in nanodiamonds can vary widely and may reach values as high as ~0.9 in favorable cases [15,16]. These reports suggest that the effective PL efficiency may be improved through sample-form optimization or surface engineering.

*Contact author: yasuyamada@chiba-u.jp

The temperature dependence of the cooling gain was further evaluated using temperature-dependent PLE measurements [24]. Figure 2(b) shows the maximum value of the cooling gain spectrum, $G_\mathrm{max}$, plotted as a function of temperature. The cooling gain increases with increasing temperature, reflecting the enhanced contribution of phonon-assisted processes at elevated temperatures. This trend suggests that anti-Stokes optical cooling is more favorable when starting from higher temperatures rather than from room temperature. However, it has also been reported that nonradiative relaxation from the triplet excited state to singlet states becomes more pronounced at elevated temperatures, leading to a reduction in PL intensity [34]. Therefore, increasing temperature does not necessarily lead to a monotonic improvement of cooling performance, and an optimal temperature range may exist.

**B. PL Lifetime**

The discussion above relies on a simple energy-balance picture that assumes a fixed NV⁻ population. In NV centers, however, optical excitation can also drive charge-state conversion between NV⁻ and NV⁰ [18,24,35-38], changing the NV⁻ cooling channel itself. We therefore examine the excitation-dependent charge-state dynamics using time-resolved PL measurements below.

We first studied the PL dynamics under excitation at 1.95 eV and 2.61 eV to quantitatively estimate the lifetimes for NV⁻ and NV⁰. As is evident from Figs. 1(b) and (c), the PL component is dominated by NV⁻ (NV⁰) under excitation at 1.95 (2.61) eV. In both cases, the decay profiles are well described by a single-exponential function, yielding lifetimes of 8.7 ns for NV⁻ and 17.9 ns for NV⁰ (see Fig. 3(a)). These values are in good agreement with those reported previously [39-41], providing reference lifetimes for separating the mixed PL dynamics under intermediate excitation conditions.

We next examined the excitation-fluence dependence of the PL dynamics under excitation at 2.31 eV, as shown in Fig. 3(b). At this excitation energy, both NV⁻ and NV⁰ can be excited, such that the detected PL dynamics reflect the combined contributions of the two charge states. With increasing excitation fluence, a long-lived component emerges. To quantify this behavior, Fig. 3(c) summarizes the excitation-fluence dependence of the effective PL lifetime for excitation photon energies in the range of 1.94–2.42 eV. The decay curves were fitted using a biexponential function, $I(t) = I_1 \exp\left(-\frac{t}{8.7\,\mathrm{ns}}\right) + I_2 \exp\left(-\frac{t}{17.9\,\mathrm{ns}}\right)$, and an effective lifetime was defined as $\tau_\mathrm{eff} = (I_1 \cdot 8.7\,\mathrm{ns} + I_2 \cdot 17.9\,\mathrm{ns})/(I_1 + I_2)$. The results show that

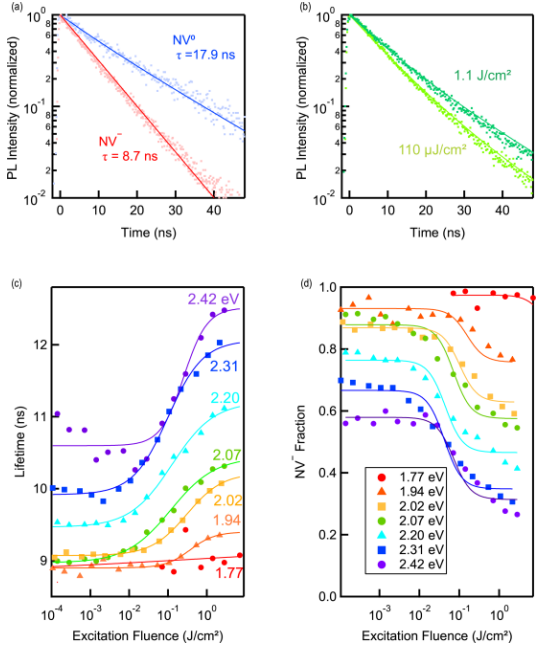

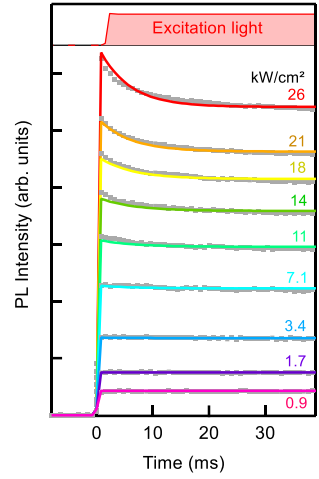

FIG. 3  (a) PL decay dynamics under excitation at 1.95 and 2.61 eV, which corresponds to $NV^-$ and $NV^0$ excitation. Solid lines represent the fitting results by exponential function. (b) PL dynamics under excitation at 2.31 eV. Excitation fluences are 110 $\mu J/cm^2$ and 1.1 $J/cm^2$, respectively. Solid curves represent the fitting results by double exponential function. (c) Effective PL lifetime under different excitation photon energy as a function of excitation fluence. (d) Fraction of the $NV^-$ component at different excitation photon energy as a function of excitation fluence. Solid curves are for guides to the eyes.

FIG. 4  PL temporal profiles measured at different excitation densities. The temporal profile of the excitation light is shown in the upper panel.

$\tau_{eff}$ increases systematically with increasing excitation fluence. In addition, the excitation fluence at which the lifetime enhancement becomes apparent shifts to higher values as the excitation photon energy is reduced.

After correcting for the spectral response of the optical system and the detector sensitivity, we obtained the excitation-fluence dependence of the $NV^-$ fraction in the total PL intensity, as shown in Fig. 3(d). In the weak-excitation regime, the $NV^-$ fraction decreases with increasing excitation photon energy, consistent with the PLE results. In the strong-excitation regime, for excitation photon energies between 1.94 and 2.42 eV, the $NV^-$ fraction decreases as the excitation fluence increases. This behavior is consistent with charge-state conversion being promoted by an increased excited-state population. Moreover, the decrease in the $NV^-$ fraction occurs at lower excitation fluences for higher excitation photon energies, which can be attributed to stronger absorption and hence a higher excited-state population even at lower excitation densities. In contrast, under 1.77-eV excitation, no significant change in the $NV^-$ fraction was observed. This can be attributed to the extremely weak absorption below the ZPL, which prevents the excited-state population from reaching a level sufficient to induce appreciable charge-state conversion.

### C. Charge-state Conversion Dynamics

To quantitatively investigate the dynamics of charge-state conversion, we focus on the temporal evolution of the PL intensity. The optically induced interconversion between $NV^0$ and $NV^-$ is expected to occur on a timescale much longer than the intrinsic radiative lifetime of the NV center. To access this timescale, the cw excitation light was modulated into a rectangular waveform using an optical chopper, and the millisecond-scale temporal evolution of the PL intensity following the onset of excitation was measured.

Figure 4 shows the results obtained under 1.95-eV excitation, which predominantly excites the $NV^-$ charge state. The PL intensity traces measured at different excitation intensities are plotted. At low excitation densities, an increase in the PL intensity is observed after the excitation is turned on, whereas at high excitation densities, a decrease in the PL intensity is observed. The relaxation behavior under strong excitation suggests a reduction in $NV^-$ population due to photoinduced charge-state conversion from $NV^-$ to $NV^0$. This excitation-density-dependent transient behavior is analyzed quantitatively in the following section using a rate-equation model.

*Contact author: yasuyamada@chiba-u.jp

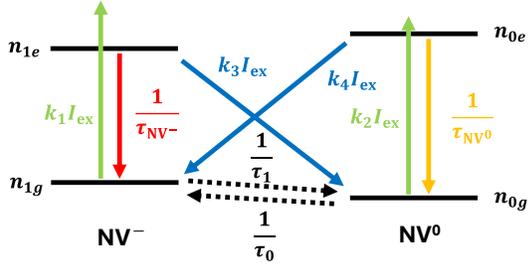

FIG. 5 Schematic diagram of the rate-equation model describing the population dynamics of the $NV^-$ and $NV^0$ charge states, including optical excitation, relaxation, and photoinduced charge-state conversion.

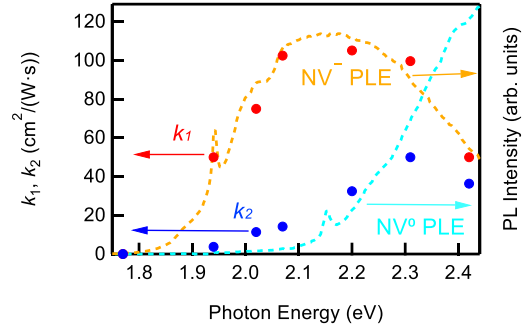

FIG. 6 Excitation-energy dependence of $k_1$ and $k_2$. The corresponding PLE spectra for $NV^-$ and $NV^0$ are also plotted.

## IV. DISCUSSION

### A. Rate-equation Analysis

The results above indicate that optical excitation drives charge-state conversion between $NV^-$ and $NV^0$, which manifests as millisecond-scale transient PL dynamics. To analyze this behavior quantitatively, we employ a minimal rate-equation model that includes the two charge states and their photoinduced interconversion. Figure 5 schematically summarizes the model. Optical pumping of $NV^-$ from the ground to the excited state (green arrow) is treated as an absorption-driven process with a rate proportional to the excitation density, $k_1 I_{ex}$, where $I_{ex}$ represents the excitation light density. The relaxation from the $NV^-$ excited state to the ground state is described with a fixed lifetime of 8.7 ns, as determined in Fig. 3(a). Similarly, for $NV^0$ we include optical pumping with rate $k_2 I_{ex}$ and relaxation with a fixed lifetime of 17.9 ns. In addition to these intra-charge-state cycles, we introduce photoinduced charge conversion through effective light-driven transitions from the excited state of $NV^-$ ($NV^0$) to the ground state of $NV^0$ ($NV^-$), with rates $k_3 I_{ex}$ and $k_4 I_{ex}$, respectively. Finally, even in the absence of optical excitation, $NV^-$ and $NV^0$ can interconvert via thermally assisted or environment-assisted electron exchange processes [35]. We describe these dark interconversion processes using two characteristic time constants, $\tau_0$ and $\tau_1$. The explicit rate equations and fitting procedures are provided in the Supplemental Material [24].

Using this rate-equation framework, we simultaneously fitted the millisecond-scale transients in Fig. 4, and determined the relevant time constants and rate coefficients. The solid curves in Fig. 4 represent the best-fit results. From the millisecond-resolved measurements, we estimate $\tau_0 = 9$ ms and $\tau_1 = 21$ ms, which characterize the dark interconversion between $NV^-$ and $NV^0$. These values are consistent with previous reports [35]. The fitting also yields the photoinduced charge-conversion coefficients $k_1 = 10.7$ cm$^2$/(W·s) and $k_3 = 1.1$ cm$^2$/(W·s).

In contrast, the excitation-energy-dependent coefficients $k_1$ and $k_2$ were obtained from the analysis of the nanosecond PL dynamics. We analyzed the $NV^-$ PL fraction shown in Fig. 3(d) using the steady-state expression Eq. (S3) of the Supplemental Material [24]. The resulting $k_1$ and $k_2$ are summarized in Fig. 6. Since $k_1$ and $k_2$ represent the effective optical pumping strength of $NV^-$ and $NV^0$, respectively, we also plot the corresponding PLE spectra on the same panel for comparison. The close agreement between the spectral shapes of $k_1$ and that of the $NV^-$ PLE spectrum, supporting the consistency of the model. The absolute value of $k_1$ obtained from the nanosecond dynamics differs by approximately a factor of four from that estimated from the millisecond dynamics. This difference likely arises because the present model is an effective description, and the extracted coefficients depend on the dynamical regime used for the analysis. While the nanosecond data reflect the initial optical pumping process, the millisecond transients more directly capture the slower charge-state redistribution relevant to the steady-state cooling behavior. Nevertheless, the central conclusion remains unchanged: photoinduced charge-state conversion limits the $NV^-$-mediated cooling cycle.

By contrast, $k_2$ shows a slight deviation from the $NV^0$ PLE spectrum. This discrepancy may reflect additional contributions under higher-energy excitation, where optical transitions other than those directly associated with $NV^0$ can also contribute to the observed response. Importantly, the uncertainty in $k_2$ has a negligible impact on the simulation results

*Contact author: yasuyamada@chiba-u.jp

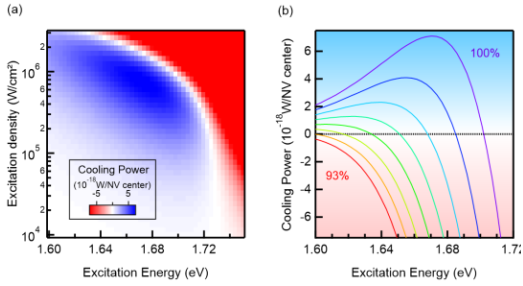

FIG. 7 (a) Calculated cooling power as a function of excitation photon energy and excitation density, with unity EQE assumed. (b) Maximum cooling power as a function of excitation photon energy for several assumed EQE values.

presented in the next section. This is because anti-Stokes excitation primarily addresses the NV$^-$ state, making the contributions of $k_2$ and $k_4$ insignificant in the cooling process.

### B. Optical Cooling Simulation

Having obtained the transition rates, we estimate the cooling performance using these experimentally constrained parameters. Figure 7(a) shows the simulated cooling power assuming a unity PL quantum efficiency, $\eta_{NV-} = 1$, for relaxation from the excited state to the ground state of NV$^-$. The cooling-power expression is written in terms of the dominant NV$^-$-mediated processes; other coefficients (e.g., $k_2$ and $k_4$) enter only through the steady-state populations (see Supplemental Material for details) [24].

In the low-excitation regime, the crossover between cooling and heating occurs near 1.73 eV, where the excitation photon energy becomes comparable to the average PL energy of NV$^-$. In the high-excitation regime, however, this crossover shifts toward lower excitation energies. This shift reflects additional heating associated with photoinduced ionization: because charge-state conversion reduces the population of the cooling-active NV$^-$ state, net cooling can only be sustained at lower excitation energies, where the cooling gain per emitted photon is larger.

Figure 7(b) shows the excitation-energy dependence of the maximum cooling power for different values of the PL quantum efficiency. A cooling window is found for quantum efficiencies of approximately 97% or higher, which is broadly consistent with the simple energy-balance estimate presented in Fig. 2.

The maximum cooling power is estimated to be $6\times10^{-18}$ W per NV center. For comparison, in perovskite quantum dots, cooling power is estimated to be ~fW per quantum dot [8]. For a representative dot diameter of 10 nm in that work, the cooling power per unit cell corresponds to ~$10^{-18}$ W, suggesting that the cooling power per NV center is comparable, on an equivalent microscopic basis.

*Contact author: yasuyamada@chiba-u.jp

A similar comparison can also be made with rare-earth-based optical coolers. Although rare-earth systems are typically limited macroscopically by their weak absorption cross sections, the cooling power per optically active ion at the fundamental limit is estimated to be of the same order of magnitude as that obtained here for a single NV center [24]. This comparison suggests that, despite the different limiting mechanisms in rare-earth ions, semiconductor nanostructures, and diamond defect centers, the attainable cooling power per active microscopic unit can be remarkably similar. By contrast, the overall cooling performance of a material is still strongly governed by macroscopic factors such as absorption strength, active-center density, and parasitic loss channels.

Under these idealized assumptions, the model indicates that substantial cooling below room temperature could in principle be possible [24]. In practice, however, the achievable temperature reduction will be strongly limited by nonunity PL efficiency, parasitic absorption, and heat leakage.

### V. CONCLUSIONS

In this work, we assessed the feasibility of anti-Stokes optical cooling in diamond NV centers. Unlike semiconductor nanostructures, where cooling performance at high excitation densities is often limited by Auger recombination, NV centers are inherently free from multiexciton-driven Auger losses. Instead, we showed that photoinduced charge-state conversion (photoionization) between NV$^-$ and NV$^0$ imposes a practical limitation on the cooling power by depleting the radiatively efficient NV$^-$ channel under excitation conditions relevant to anti-Stokes operation. Within our experimentally constrained effective model, the resulting cooling power per center is comparable, on a microscopic basis, to values discussed not only for recently demonstrated perovskite quantum-dot refrigeration but also for rare-earth-based optical cooling at the single-ion level. This implies that, although the dominant limiting mechanisms differ substantially among these material platforms, the attainable cooling power per active microscopic unit can be of a similar order of magnitude. Net optical refrigeration could be achievable in diamond defect systems, provided that sufficiently high PL efficiency can be maintained under operating conditions. At the same time, the external PL efficiency of typical NV ensembles remains far from unity, motivating the exploration of alternative color centers with higher radiative yield, as potentially more favorable platforms for optical cooling.

Our analysis also indicates that the self-limiting behavior arising from photoionization can, in principle, be mitigated by accelerating the recovery from $NV^0$ to $NV^-$ (i.e., by reducing $\tau_0$), since rapid recharging would help maintain the population of the cooling-active charge state. For example, phosphorus doping can stabilize the $NV^-$ charge state, enabling nearly complete conversion of NV centers into $NV^-$ [42]. If parasitic absorption introduced by dopants can be kept sufficiently low, such charge-state engineering is expected to enhance the attainable cooling power [24] and improve the prospects for steady-state anti-Stokes optical refrigeration.

In addition, new optical-cooling schemes based on quantum coherent control have been theoretically proposed for diamond NV centers [43-45]. Our results provide an important experimental basis for evaluating the feasibility of such approaches under realistic operating conditions. Overall, the present study provides quantitative guidance toward realizing optical cooling in diamond defect systems by identifying charge-state conversion as a key bottleneck and by clarifying experimentally grounded pathways to overcome it.

ACKNOWLEDGMENTS

This work was supported by Asahi glass foundation, JSPS KAKENHI (JP25K01649), and JSPS Program for Forming Japan's Peak Research Universities（JPJS00420230002）.

*Contact author: yasuyamada@chiba-u.jp

*Contact author: yasuyamada@chiba-u.jp